\documentclass[conference]{IEEEtran}

\ifCLASSINFOpdf
   \usepackage[pdftex]{graphicx}
\else
\fi
\usepackage[cmex10]{amsmath}
\usepackage{array}

\usepackage{mdwmath}
\usepackage{mdwtab}

\usepackage{fixltx2e}

\usepackage{url}

\usepackage{esdiff}
\usepackage{mathtools}
\usepackage{array}
\usepackage{placeins}
\newcommand{\christoffel}[3]{\ensuremath{\Gamma^{#1}_{#2#3}}}

\begin{document}

\title{On geodesic deviation in Schwarzschild spacetime}

\author{\IEEEauthorblockN{Dennis Philipp,
Volker Perlick, 
Claus L\"ammerzahl and Kaustubh Deshpande}
\IEEEauthorblockA{ZARM - Center of Applied Space Technology and Microgravity\\
University of Bremen \\
28359 Bremen, Germany\\\\ Email: dennis.philipp@zarm.uni-bremen.de}}

\maketitle

\begin{abstract}
For metrology, geodesy and gravimetry in space, satellite based instruments and measurement techniques are used and the orbits of the satellites as well as possible deviations between nearby ones are of central interest. The measurement of this deviation itself gives insight into the underlying structure of the spacetime geometry, which is curved and therefore described by the theory of general relativity (GR).
In the context of GR, the deviation of nearby geodesics can be described by the Jacobi equation that is a result of linearizing the geodesic equation around a known reference geodesic with respect to the deviation vector and the relative velocity. We review the derivation of this Jacobi equation and restrict ourselves to the simple case of the spacetime outside a spherically symmetric mass distribution and circular reference geodesics to find solutions by projecting the Jacobi equation on a parallel propagated tetrad as done by Fuchs \cite{Fuchs1990}. Using his results, we construct solutions of the Jacobi equation for different physical initial scenarios inspired by satellite gravimetry missions and give a set of parameter together with their precise impact on satellite orbit deviation.
We further consider the Newtonian analog and construct the full solution, that exhibits a similar structure, within this theory.
\end{abstract}

\IEEEpeerreviewmaketitle


\section{Introduction}
When satellites follow freely falling orbits around a central massive object like the earth, their worldlines, i.e., their paths through space and time must be described by the geodesic equation together with given initial conditions. While for some (past) space missions the Newtonian theory of gravity might be sufficient, modern and future mission scenarios certainly need relativistic effects to be taken into account. Thus, the precise description in terms of Einstein's theory of gravity, GR, becomes necessary even beyond usual Post-Newtonian (PN) approximations.
In this work we will describe the geodesic deviation, with satellite missions in mind, in the full context of GR.

As an example, the GFZ--NASA mission GRACE-Follow-On \cite{Grace-FO, Dehne2009} consists of two satellites which are able to measure the change in their relative distance (about 100 km) with an accuracy of the order of 10 nm. For the orbital motion of the two satellites we can imagine different configurations:
\begin{itemize}
\item[i)] Tilt the orbital plane of one satellite with respect to the other, but keep the constants of motion (in magnitude) the same. This orbital configuration is called a pendulum orbit.
\item[ii)] For a second configuration, the orbital plane is the same, but the energy and angular momentum are slightly different and the result is the so-called cartwheel orbit. 
\item[iii)] A more general possibility that allows to change the orbital plane as well as the constants of motion is the helical orbit configuration.
\end{itemize}

For the precise measurements of inter-satellite distances in these orbit configurations the general relativistic effects must be investigated and their impact on observables taken into account for a given accuracy level. In this work we will focus on such orbital configurations and refer to one satellite as the reference object moving on the reference geodesic. The orbit of the second satellite will then be modeled by means of the geodesic deviation equation that describes how nearby geodesics deviate from each other due to the geometry of spacetime and given initial conditions. We will first describe the situation in standard Newtonian gravity and then turn to the full theory of GR to uncover relativistic modifications.
\section{Geodesic deviation in Newtonian gravity}
In the Newtonian theory of Gravity we have as central equations
\begin{subequations}
\begin{align}
\Delta U(\vec{r}) &= -4\pi G \rho(\vec{r}) \, , \label{eq:NewtonFieldEq}\\
\diff[2]{\vec{r}}{t} &= - \nabla U(\vec{r}) \, , \label{eq:NeqtonEOM}
\end{align}
\end{subequations}
where the first of them is the field equation that relates the Newtonian gravitational potential $U$ to the mass density $\rho$ and introduces Newton's gravitational constant as a factor of proportionality. Outside a spherically symmetric object we obtain the gravitational potential
\begin{align}
U(r) = \dfrac{GM}{r} \label{eq:NewtonPotential}
\end{align}
as a solutions of \eqref{eq:NewtonFieldEq}. The second equation \eqref{eq:NeqtonEOM} is the equation of motion and describes how test particles move in the gravitational potential given by $U$. The second derivative of the position vector is taken w.r. to universal time $t$ that exists in Newtonian gravity. We can rewrite the second equation using index notation and obtain
\begin{align}
\ddot{x}^a = - \partial^a U(x). \label{eq:NeqtonEOM2}
\end{align}
The coordinates $x^a$ are just the usual Cartesian coordinates and the overdot denotes derivatives w.r. to the Newtonian time coordinate. The argument $x$ in the potential denotes the position of the test particle. Here and in the following, Latin indices $a,b,...$ take values $1,2,3$. Now, we assume a situation as shown in Fig. \ref{fig:deviation}, i.e., we have a reference curve $X^a (t)$ that fulfills \eqref{eq:NeqtonEOM2}. Thereupon, we consider a second curve $x^a(t) = X^a(t) + \eta^a(t)$ and introduce the deviation vector $\eta^a$. Thus, we have
\begin{align}
\ddot{x}^a = \ddot{X}^a + \ddot{\eta}^a = - \partial^a U(x) = - \partial^a U(X + \eta).
\end{align}
Now, we linearize the right hand side w.r. to the deviation
\begin{align}
U(x) = U(X+\eta) = U(X) + \eta^a \partial_a U(X) + \mathcal{O}(\eta^2) \, ,
\end{align}
and obtain finally the deviation equation in Newtonian gravity
\begin{align}
\ddot{\eta}^a = - \eta^b \partial^a \partial_b U(X) \, . \label{eq:NewtonDeviationEq}
\end{align}
From \eqref{eq:NewtonDeviationEq} we clearly see that second derivatives of the Newtonian potential cause non-linear deviations. If either $\partial_a U = 0$ (homogeneous gravitational field) or $U \equiv 0$ (no gravitational field) the deviation equation has the solution
\begin{align}
\eta^a (t) = C_1 t + C_2
\end{align}
and the deviation vector grows only linearly in time. In the following, we use the Newtonian gravitational potential \eqref{eq:NewtonPotential} outside a spherically symmetric mass distribution and we change to usual spherical coordinates $(x,y,z) = (r\sin\vartheta \cos\varphi, r\sin\vartheta \sin\varphi, r \cos\vartheta)$. We further specialize the reference geodesic $(R(t), \Theta(t), \Phi(t))$ to be a circular orbit in the equatorial plane. Thus, we have
\begin{subequations}
\begin{align}
R(t) &= R_0 = \text{const.} \, , \\
\Theta(t) &= \Theta_0 \equiv \pi/2 \, , \\
\Phi(t) &= \sqrt{\dfrac{GM}{r^3}} t =: \omega_0 t \, ,
\end{align}
\end{subequations}
where the motion is oscillating with the Keplerian period $T_0 = 2\pi / \omega_0 = 2\pi \sqrt{r^3/(GM)}$.
For this situation the deviation equation \eqref{eq:NewtonDeviationEq} reduces to three differential equations for the components of the deviation vector $(\eta^r, \eta^\vartheta, \eta^\varphi)$, see \cite{Greenberg1974}, 
\begin{subequations}
\begin{align}
\diff[2]{\eta^\vartheta}{t} + \omega^2_0 \eta^\vartheta = 0 \, , \\
\diff[2]{\eta^r}{t} - 2 \omega_0 \diff{\eta^\varphi}{t} - 3\omega_0 \eta^r = 0 \, , \\
\diff[2]{\eta^\varphi}{t} + 2 \omega_0 \diff{\eta^r}{t} = 0 \, .
\end{align}
\end{subequations}
The first equation decouples from the other two and the solution is given by 
\begin{align}
\eta^\vartheta(t) = C_5 \cos \omega_0 t + C_6 \sin \omega_0 t \, ,
\end{align}
which is an oscillation with the Keplerian period $T_\vartheta = T_0 = 2\pi / w_0$. In \cite{Greenberg1974} the author derived only the oscillating solutions for the remaining two equations. However, we extend this and give the general solution that yields
\begin{subequations}
\begin{align}
\eta^r (t) &= C_1 + C_2 \sin \omega_0 t + C_3 \cos \omega_0 t \, , \\
\eta^\varphi (t) &= -\dfrac{3}{2} \omega_0 C_1 t + 2 (C_2 \cos \omega_0 t - C_3 \sin \omega_0 t) +  C_4 \, .
\end{align}
\end{subequations}
There are several possibilities to perturb the circular reference geodesic. One can incline the orbital plane, add a constant radius or cause an eccentricity in the motion. All these effects are due to the choice of the six parameter $C_i$. We must have precisely six free parameter to set the initial position and velocity as starting conditions. The meaning of these parameter and their impact on the perturbed orbit will be given in section \ref{sec:parameter} in one go with those for the GR results.
\section{Geodesic deviation equation in GR}
\begin{figure}
\centering
\includegraphics[width=0.3\textwidth]{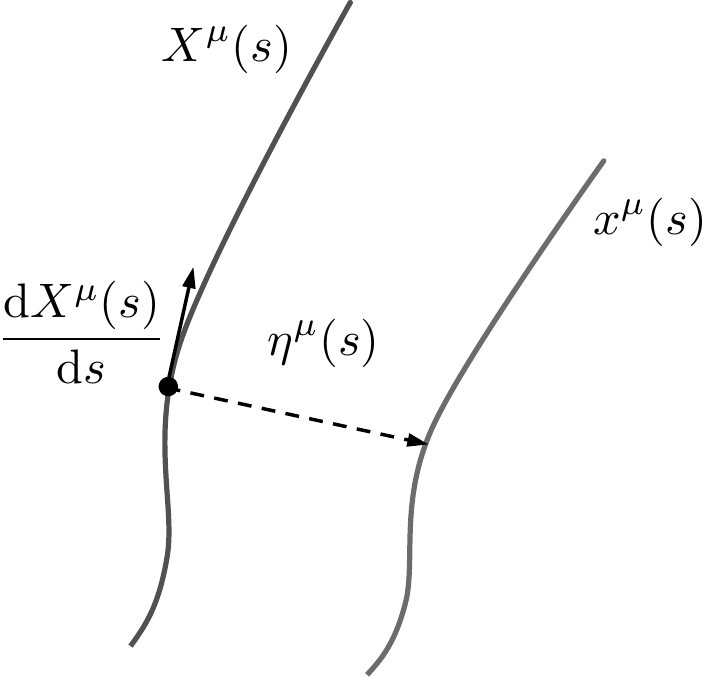}
\caption{\label{fig:deviation}The deviation of two nearby geodesics $X^\mu(s)$ and $x^\mu(s) = X^\mu(s) + \eta^\mu(s)$. Note: the deviation vector is always defined orthogonal (as measured with the metric $g$) on the four-velocity $\mathrm{d}X^\mu(s)/\mathrm{d}s$ of the reference geodesic.}
\end{figure}

In the spirit of general relativity we model the spacetime geometry, i.e., the universe we live in (or at least the relevant part for our model) using a pseudo-Riemannian metric tensor
\begin{align}
g = g_{\mu\nu} dx^\mu dx^\nu
\end{align}
on a four dimensional manifold $\mathcal{M}$. The set $\left\lbrace \mathcal{M},g \right\rbrace$ describes the four-dimensional spacetime and local coordinates on $\mathcal{M}$ are $x^\mu \, , \mu=0,1,2,3$.
We have an affine connection $\nabla$, which is fully defined by the Christoffel symbols
\begin{align}
\nabla_{\partial_\nu} \partial_\mu = \christoffel{\sigma}{\mu}{\nu} \partial_\sigma \, .
\end{align}
On a pseudo-Riemannian manifold we can specialize $\nabla$ to be the Levi-Civita connection and we get
\begin{align}
\christoffel{\sigma}{\mu}{\nu} = \dfrac{1}{2} g^{\sigma \lambda} (\partial_\nu g_{\mu \lambda} + \partial_\mu g_{\nu \lambda} - \partial_\lambda g_{\mu \nu}) \, .
\end{align}
Then, the geodesic equation that describes the motion of freely falling particles reads
\begin{align}
\diff[2]{x^\mu}{s} + \christoffel{\mu}{\nu}{\sigma}(x) \diff{x^\nu}{s} \diff{x^\sigma}{s} = 0 \label{eq:geodesicEquation}
\end{align}
and gives as solution curves the geodesics $x^\mu (s)$ of the spacetime. The affine parameter $s$ along such a geodesic can be interpreted as proper time and is, thus, related to the reading of a clock that is transported along the geodesic. Now, we fix a certain (known) geodesic $(X^\mu (s)) = (X^0(s),X^1(s),X^2(s),X^3(s))$ and this curve will be called the reference geodesic in the following. To consider a neighboring geodesic $x^\mu (s)$ in a given coordinate system we make the ansatz
\begin{align}
x^\mu (s) = X^\mu (s) + \eta^\mu (s) \label{eq:deviationVector}
\end{align}
and define, thereupon, the deviation vector $\eta^\mu (s)$ that connects both geodesics. 
We assume, as sketched in Fig. \ref{fig:deviation}, the four velocity of the reference geodesic and the deviation vector to be always orthogonal to each other,
\begin{align}
g_{\mu\nu} \eta^\mu \diff{X^\nu}{s} = 0 \, .
\end{align}
Inserting \eqref{eq:deviationVector} into the geodesic equation \eqref{eq:geodesicEquation} gives a second order differential equation that is quadratic in the deviation vector. When this deviation vector is assumed to be very small, we can linearize with respect to the deviation vector itself $\eta^\mu (s)$ and its derivative $\mathrm{d} \eta^\mu (s)/\mathrm{d}s$ if we further assume small relative velocities. Thus, we obtain the so called standard Jacobi equation
\begin{align}
\dfrac{\mathrm{D}^2 \eta^\mu (s)}{\mathrm{d} s^2} = - R^{\mu}_{~ \tau\nu\sigma}(X) \, 
\eta^\nu \diff{X^\tau (s)}{s} \diff{X^\sigma (s)}{s} \, , \label{eq:JacobiEqn}
\end{align}
where the covariant derivative $\mathrm{D}/\mathrm{d}s$ and the curvature tensor components $R^{\mu}_{~ \tau\nu\sigma}$ are given by
\begin{subequations}
\begin{align}
\dfrac{\mathrm{D} \, \eta^\mu (s)}{\mathrm{d} s} = \diff{\eta^\mu (s)}{s} + \christoffel{\mu}{\nu}{\sigma}(X) \, \eta^\nu \diff{X^\sigma}{s} \, , \\[1em]
R^{\mu}_{~ \tau\nu\sigma}(X) = \partial_\nu \christoffel{\mu}{\tau}{\sigma} - \partial_\tau \christoffel{\mu}{\nu}{\sigma} + \christoffel{\mu}{\nu}{\lambda} \christoffel{\lambda}{\tau}{\sigma}-\christoffel{\mu}{\tau}{\lambda}\christoffel{\lambda}{\nu}{\sigma} \, .
\end{align}
\end{subequations}
In \eqref{eq:JacobiEqn} we can clearly see that the curvature of spacetime induces a possible non-linear deviation between two neighboring geodesics. We have seen in the Newtonian case before that the same effect of non-linear deviation was caused by non-vanishing second derivatives of the Newtonian gravitational potential and we have, thus, an intuitive understanding of the role of the curvature tensor; The Riemann curvature tensor includes second derivatives of the metric as well.
A somewhat detailed discussion of the equation of geodesic deviation can be found in standard textbooks on general relativity like \cite{MTW1973} and \cite{HawkingEllis1973}.
It should be mentioned that if we do not linearize w.r. to the relative velocity $\mathrm{d}\eta(s)/\mathrm{d}s$ a generalized version of the Jacobi equation is obtained, see, e.g., \cite{Hodgkinson1972,Ciufolini1986,Mashhoon2002,Perlick2008}.

In the following we specialize the spacetime to be spherically symmetric and static, described by the metric
\begin{align}
g = -A(r) dt^2 + B(r) dr^2 + r^2 (d\vartheta^2 + \sin^2 \vartheta d\vartheta^2) \label{eq:MetricSphericallySym}
\end{align}
and introduce the coordinates $x^0 = t, x^1 = r, x^2 = \vartheta, x^3 = \varphi$. The angles $\vartheta$ and $\varphi$ are the usual polar and azimuthal angles as in spherical polar coordinates and the radial coordinate $r$ is defined such that circles at a distance $r$ have circumference $2\pi r$. In these coordinates the reference geodesic is given by
\begin{align}
&X^0(s) = T(s), \, X^1(s) = R(s), \notag \\
&X^2(s) = \Theta(s), \, X^3(s) = \Phi(s) \, .
\end{align}
For a metric of the form \eqref{eq:MetricSphericallySym} we can always, without loss of generality, assume that the reference geodesic is fixed in the equatorial plane by choosing the coordinate system to match this condition. This is due to the spherical symmetry of the spacetime exhibited in \eqref{eq:MetricSphericallySym} and implies i) $\Theta(s) = \pi/2 = \text{const.}$ and ii) $\mathrm{d}\Theta(s)/ \mathrm{d}s = 0$. Since we wish to actually describe the motion of satellites, we must restrict to timelike geodesics, which describe the motion of massive test particles at subluminal speed. For such geodesics we can identify the parameter $s$ along the reference geodesic with the proper time according to the normalization of the four velocity
\begin{align}
-1 = g_{\mu\nu} \diff{X^\mu (s)}{s}\diff{X^\nu (s)}{s} \, .
\end{align}
Here, we use natural units in which Newtons gravitational constant $G$ and the speed of light $c$ take the value $c=G=1$.
For such timelike geodesics in the metric \eqref{eq:MetricSphericallySym} we obtain constants of motion that correspond to the conservation of energy $E$ and angular momentum $L$, see for example \cite{Fuchs1977}. These constants can be derived using the Euler Lagrange equations
\begin{align}
\diff{}{s} \dfrac{\partial \mathcal{L}}{\partial \dot{x}^\mu} - \dfrac{\partial \mathcal{L}}{\partial x^\mu} = 0
\end{align}
for the Lagrangian 
\begin{align}
\mathcal{L} = \dfrac{1}{2} g_{\mu\nu} \dot{x}^\mu \dot{x}^\nu.
\end{align}
Since the metric \eqref{eq:MetricSphericallySym} does neither depend on the time coordinate $t$ nor on the angle $\phi$ we get for the reference geodesic
\begin{subequations}
\begin{align}
E &:= A(R(s)) \, \dot{T}(s) = \text{const.} \, ,\\
L &:= R(s)^2 \, \dot{\Phi}(s) = \text{const.} \, ,
\end{align}
\end{subequations}
where the overdot denotes the derivative with respect to the proper time $s$. 
The general solution of the geodesic deviation equation \eqref{eq:JacobiEqn} in the spacetime \eqref{eq:MetricSphericallySym} was given by Fuchs \cite{Fuchs1983} in terms of first integrals of the Jacobi equation. This solution is, however, not applicable for circular reference geodesics since in this case the condition $\mathrm{d}R(s)/\mathrm{d}s = 0$ holds and causes singularities in the equations in this reference. Shirokov derived periodic solutions for the Schwarzschild spacetime and circular reference geodesics \cite{Shirokov1973}.
One different possibility to solve the geodesic deviation equation for the case of circular reference geodesics in the spacetime \eqref{eq:MetricSphericallySym} is to refer it to a parallel propagated tetrad and solve the resulting differential equation in this reference system \cite{Fuchs1990}. We will use the results of this method here.

The simplest model that we could use to describe the motion of satellites in orbit around the earth is to specialize to the case of Schwarzschild spacetime
\begin{align}
A(r) = \left( 1 - \dfrac{2M}{r} \right) \, , \quad B(r) = A(r)^{-1}
\end{align}
and the circular reference geodesic is essentially determined through its radius $R_0$ and the mass $M$ of the earth that enters the metric coefficients
\begin{subequations}
\begin{align}
R(s) &\equiv R_0 = \text{const.} \\
\Theta(s) &\equiv \dfrac{\pi}{2} = \text{const.}  \\
\Phi(s) &= \dot{\Phi} \, s = \dfrac{L}{R_0^2} \, s =: \omega s \\
T(s) &= \dot{T} \, s = \dfrac{E}{A(R_0)} \, s = \dfrac{E}{1-2M/R_0} \, s \, .
\end{align}
\end{subequations}
We should mention that the mass of the earth in the units that we use is about $M \approx 0.5\,$cm. After some lengthy calculations one arrives at the solution of the Jacobi equation using the result of Fuchs in \cite{Fuchs1990}
\begin{subequations}
\label{eq:solutionCircRef}
\begin{align}
\eta^0(s) &= \dfrac{L E}{(1-2M/R_0) \sqrt{L^2+R_0^2}} \, f(s) \\
\eta^1(s) &= \dfrac{E R_0}{\sqrt{L^2+R_0^2}} \, g(s) \\
\eta^2(s) &= \dfrac{C_5}{R_0} \cos \omega s + \dfrac{C_6}{R_0} \sin \omega s\\
\eta^3(s) &= \dfrac{\sqrt{L^2+R_0^2}}{R_0^2} \, f(s) \, ,
\end{align}
\end{subequations}
where the two proper time dependent function $f(s)$ and $g(s)$ are given by
\begin{subequations}
\begin{align}
f(s) &= C_1 \left(1-\dfrac{4R_3}{k^2} \right)s + \sqrt{\dfrac{4R_3}{k^2}} (C_2 \cos ks - C_3 \sin ks) \notag \\
&+ C_4 \, , \\[1em]
g(s) &= \dfrac{C_1}{k} \sqrt{\dfrac{4R_3}{k^2}} + C_2 \sin ks + C_3 \cos ks \, .
\end{align}
\end{subequations}
The constants of motion $E$ and $L$ as well as the remaining quantities $k$ and $R_3$ are uniquely defined by the radius $R_0$ of the circular reference geodesic
\begin{subequations}
\begin{align}
\label{eq:constants}
R_{3} &= \dfrac{M}{R_0^3} \, , \\
k^2 &= \dfrac{M(R_0-6M)}{R_0^3(R_0-3M)} \, , \\
E^2 &= \dfrac{(R_0-2M)^2}{R_0(R_0-3M)} \, , \\
L^2 &= \dfrac{MR_0^2}{R_0-3M} \, .
\end{align}
\end{subequations}
All the other constants $C_i, i=1..6$ can be used to model different initial conditions.
\section{Results}
\subsection{ \label{sec:parameter} The initial conditions}
To describe different orbital configurations we have to examine the meaning of the constants $C_i$. A proper way to do this is to investigate the impact of each constant separately, i.e., having only one of them unequal to zero.
The parameter $C_i$ then yield the following meaning:
\begin{itemize}
\item[$C_1$:] In the case that only $C_1 \neq 0$ we obtain again a circular orbit in the equatorial plane with radius
\begin{align*}
x^1(s) = r(s) & = R(s) + \eta^1(s) \\
&= R_0 + \dfrac{E R_0}{\sqrt{L^2+R_0^2}} C_1 \dfrac{2\sqrt{R_3}}{k^{2}} \\
&= R_0 + \delta r
\end{align*}
We have to ensure that the perturbation $\delta r$ is small in comparison to the reference radius $R_0$ by means of choosing $C_1$ in a proper way.
\item[$C_{2,3}$:] These two constants will cause an elliptical motion in the $r-\phi$ plane, e.g., if only $C_3 \neq 0$ we get a radial and azimuthal motion of the form 
\begin{align*}
r(s) &= R_0 + \dfrac{E R_0}{\sqrt{L^2+R_0^2}} \, C_3 \cos ks  = R_0 + \delta r \cos ks \\
\phi(s) &= \Phi(s) - \dfrac{\sqrt{L^2+R_0^2}}{R_0^2} \dfrac{2\sqrt{R_3}}{k} C_3 \sin ks \\
&= \omega s - \delta \phi \sin ks
\end{align*}
For a perturbed motion of this kind, the eccentricity $e$ and the semi major axis $a$ are then given by
\begin{align*}
e = \dfrac{\delta r}{R_0} \, , \quad a = R_0 \, .
\end{align*}
\item[$C_4$:] The constant $C_4$ simply describes an offset in the azimuthal and temporal components, e.g., when only $C_4 \neq 0$ we have a motion on the same circle as the reference geodesic but with a constant azimuthal separation
\begin{align*}
\phi(s) = \Phi(s) + \dfrac{\sqrt{L^2+R_0^2}}{R_0^2} \, C_4 = \omega s + \delta \phi
\end{align*} 
\item[$C_{5,6}$:] These two constants incline the orbit w.r. to the reference geodesic, e.g., if only $C_5 \neq 0$ one obtains a circular orbit with radius $R_0$ but with a polar angle
\begin{align*}
\theta(s) &= \Theta(s) + \dfrac{C_5 \cos \omega s}{R_0} \\
&= \dfrac{\pi}{2} + \delta \theta \cos \omega s
\end{align*}
\end{itemize}
By choosing more than one constant unequal to zero at the same time, we get combinations of the described effects. We have named the constants $C_i$ in the Newtonian calculation in the same way, such that the before mentioned effects of these constants are qualitatively the same.

In the following we examine two examples for possible orbit deviations in some detail and give a visual representation in Fig. (\ref{fig:pendulum}) and (\ref{fig:EccOrbit}).
\subsection{Pendulum orbits}
\begin{figure*}[h]
\begin{center}
\includegraphics[width=0.3\textwidth]{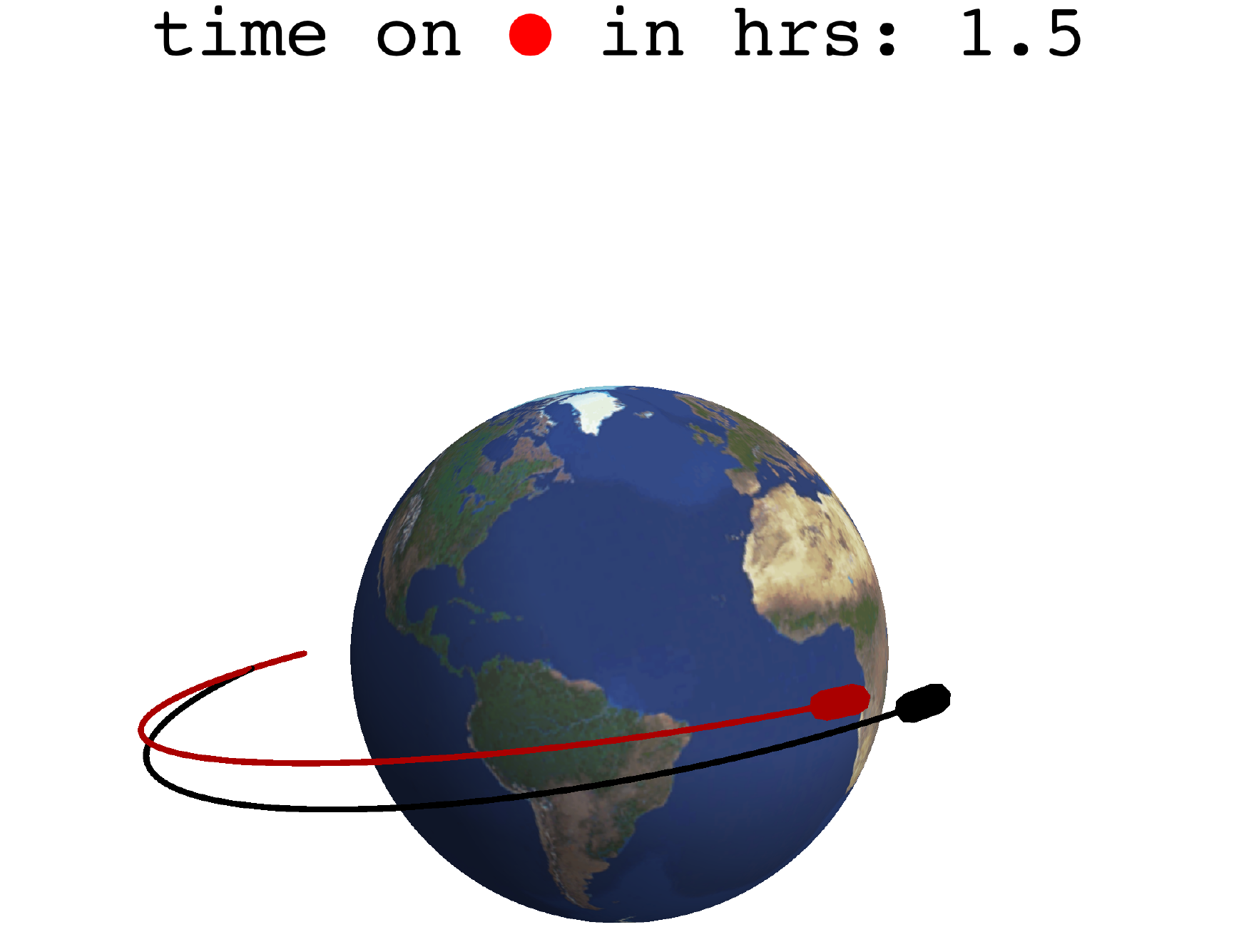}
\includegraphics[width=0.3\textwidth]{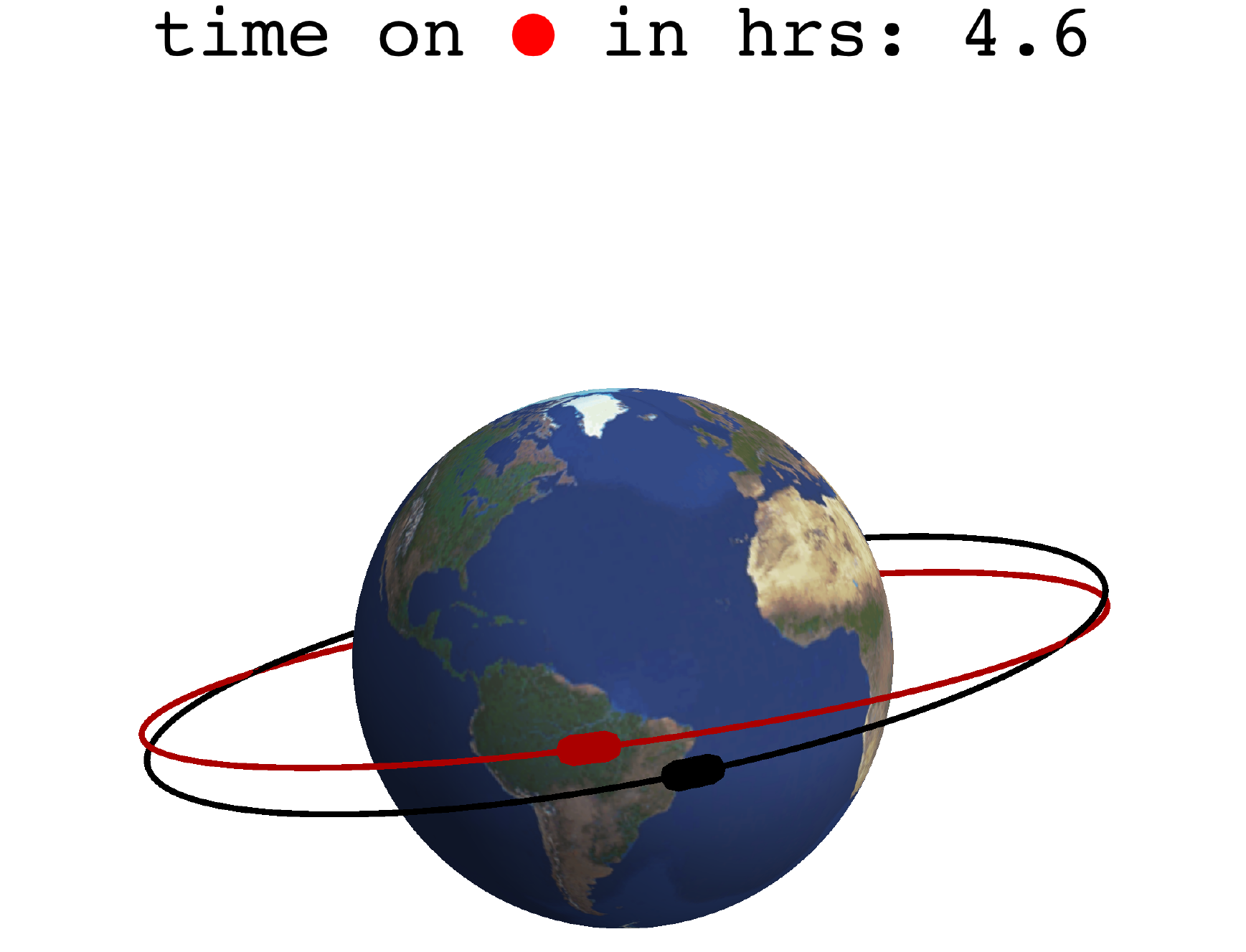}
\includegraphics[width=0.3\textwidth]{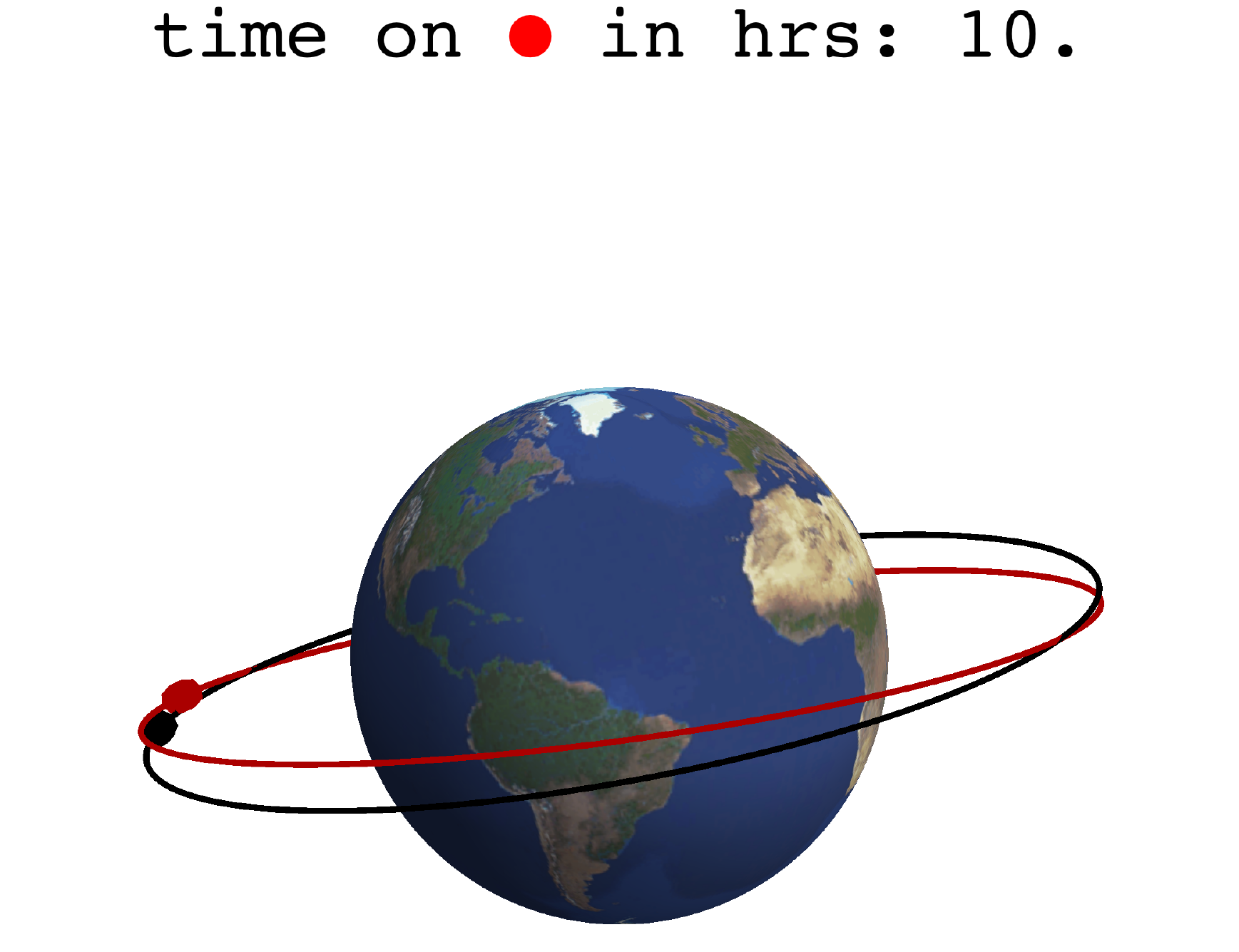}
\caption{\label{fig:pendulum} A pendulum orbit: the deviating geodesic (black) is now inclined but circular again with the same radius as the reference geodesic (red) to keep the energy and angular momentum the same since these are purely determined by the radius of the circular motion. We included an azimuth angle offset via $C_4$ to prevent both objects from colliding.}
\end{center}
\end{figure*}

One possible application of the above results is to model a pendulum orbit, where the orbital planes of two satellites are inclined w.r. to each other but we keep the constants of motion the same. For a circular reference geodesic this means that the radius of the second satellite's orbit is the same and the orbit is therefore circular again. To achieve this, we have chosen $C_1=C_2=C_3=0$. The precise choice of $C_5,C_6$ determines the line of nodes and we can include an azimuthal offset using $C_4$ to prevent both satellites from colliding. A result of this kind is shown in Fig.  (\ref{fig:pendulum}) for different values of the elapsed proper time. The perturbed motion is again circular with
\begin{subequations}
\begin{align}
r(s) &= R_0 \, ,\\
\theta(s) &= \pi/2 + C_5/R_0 \cos \omega s + C_6/R_0 \sin \omega s \, , \\
\phi(s) &= \omega s + C_4 \, .
\end{align}
\end{subequations}
This result holds equally well in the Newtonian case, there we can also get a perturbed orbit like this and the result shown in Fig. (\ref{fig:pendulum}) is the same.
\subsection{Cartwheel orbits}
\begin{figure*}[h]
\centering
\includegraphics[width=0.3\textwidth]{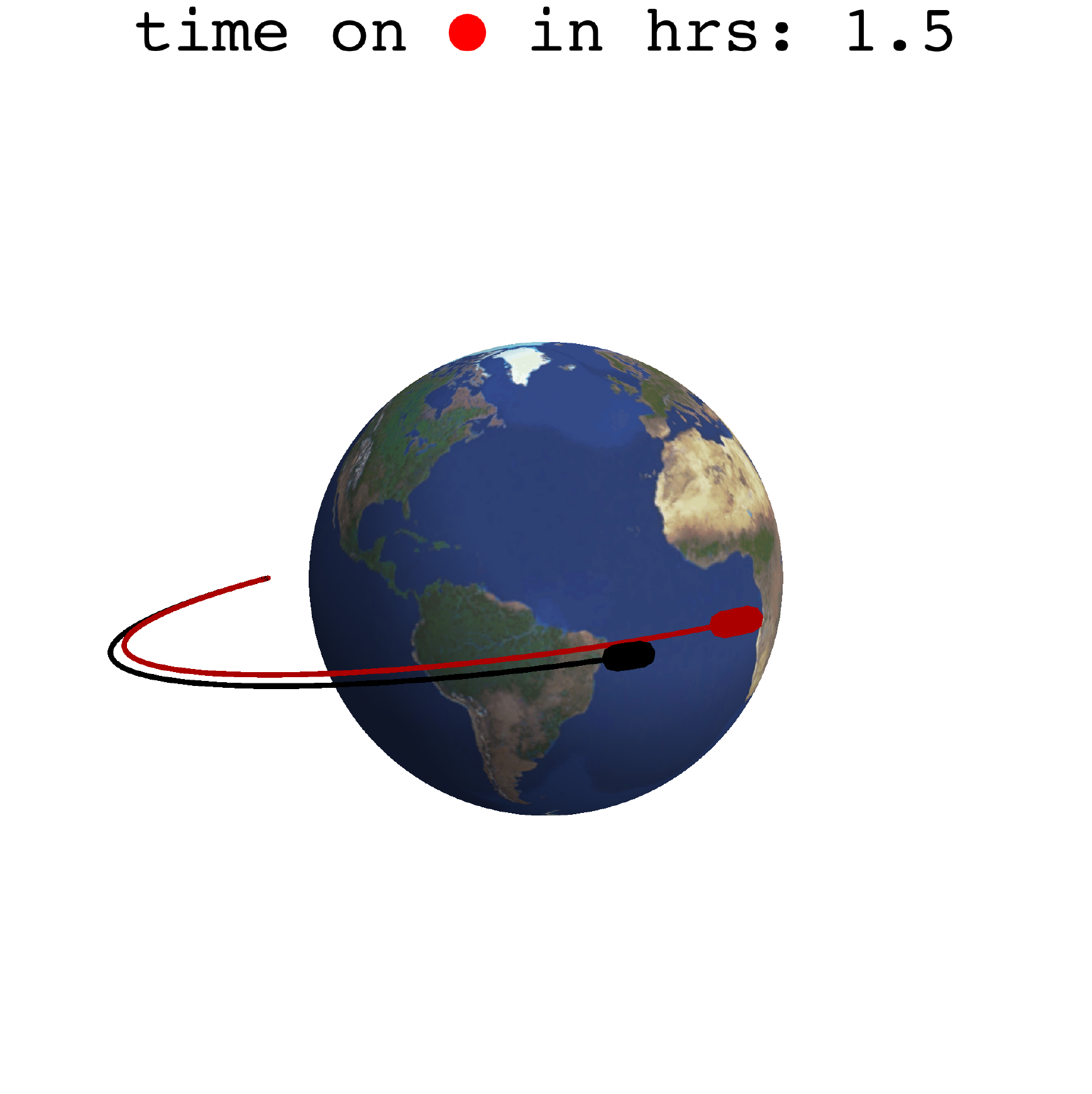}
\includegraphics[width=0.3\textwidth]{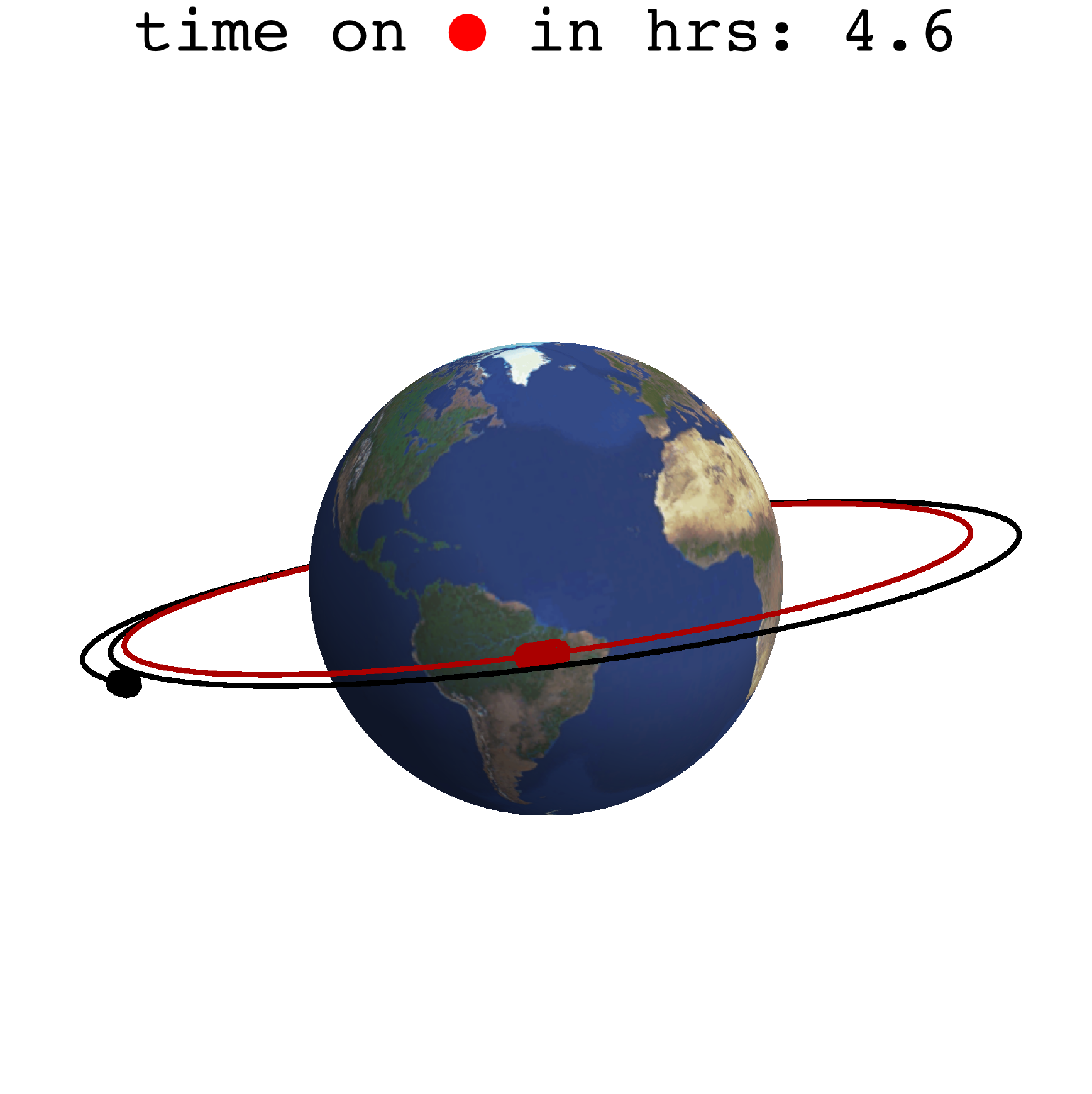}
\includegraphics[width=0.3\textwidth]{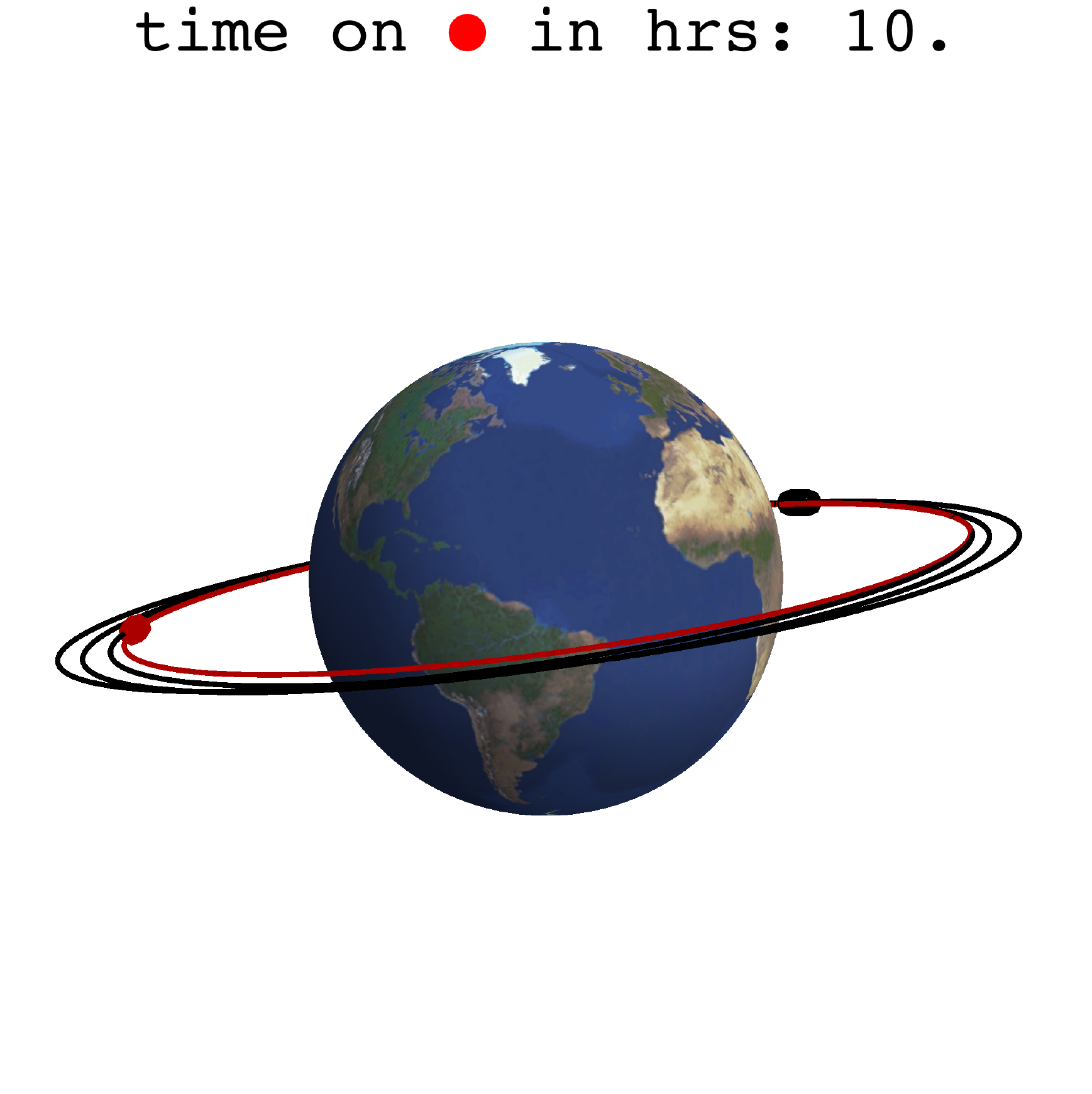}
\includegraphics[width=0.3\textwidth]{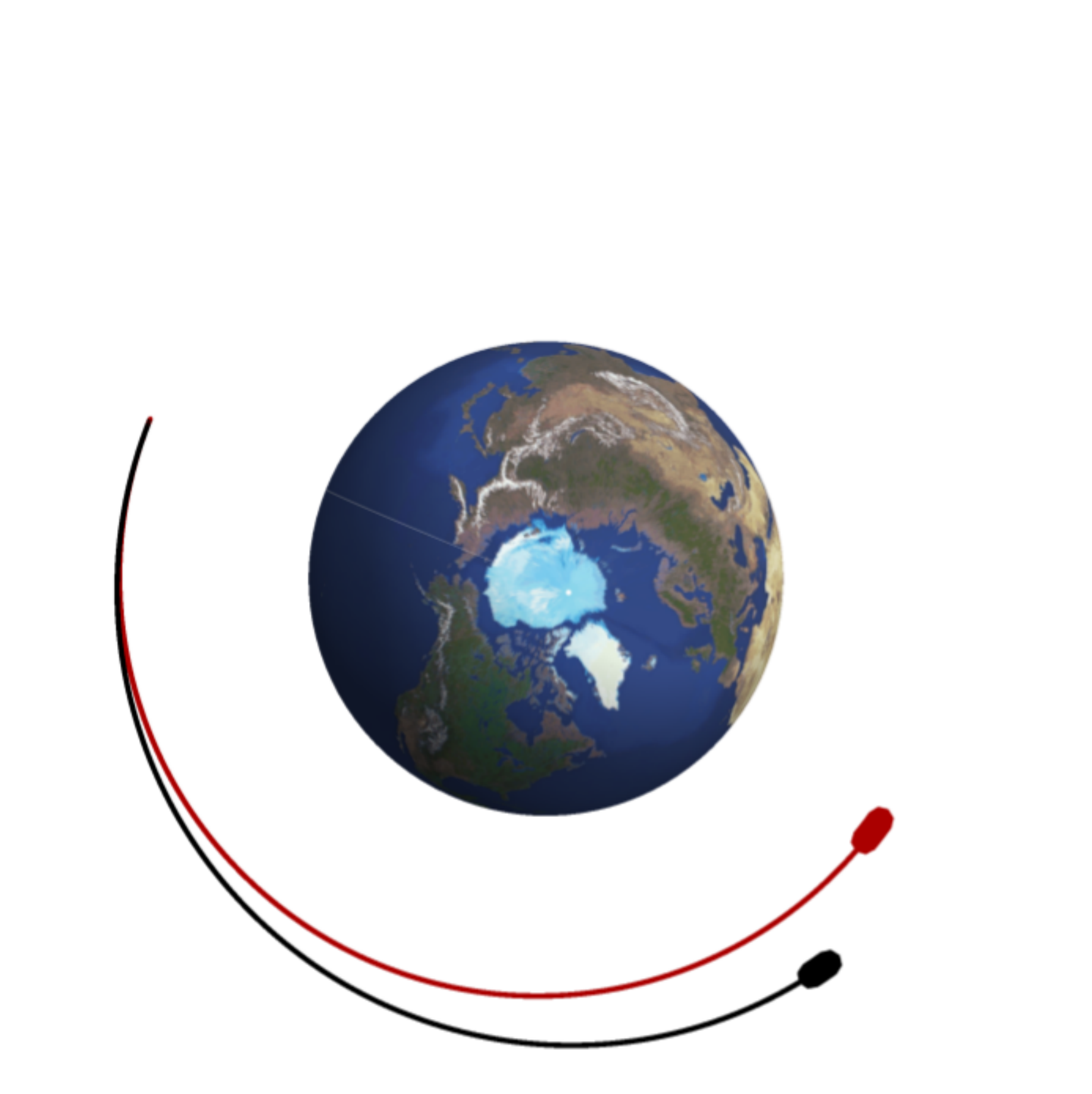}
\includegraphics[width=0.3\textwidth]{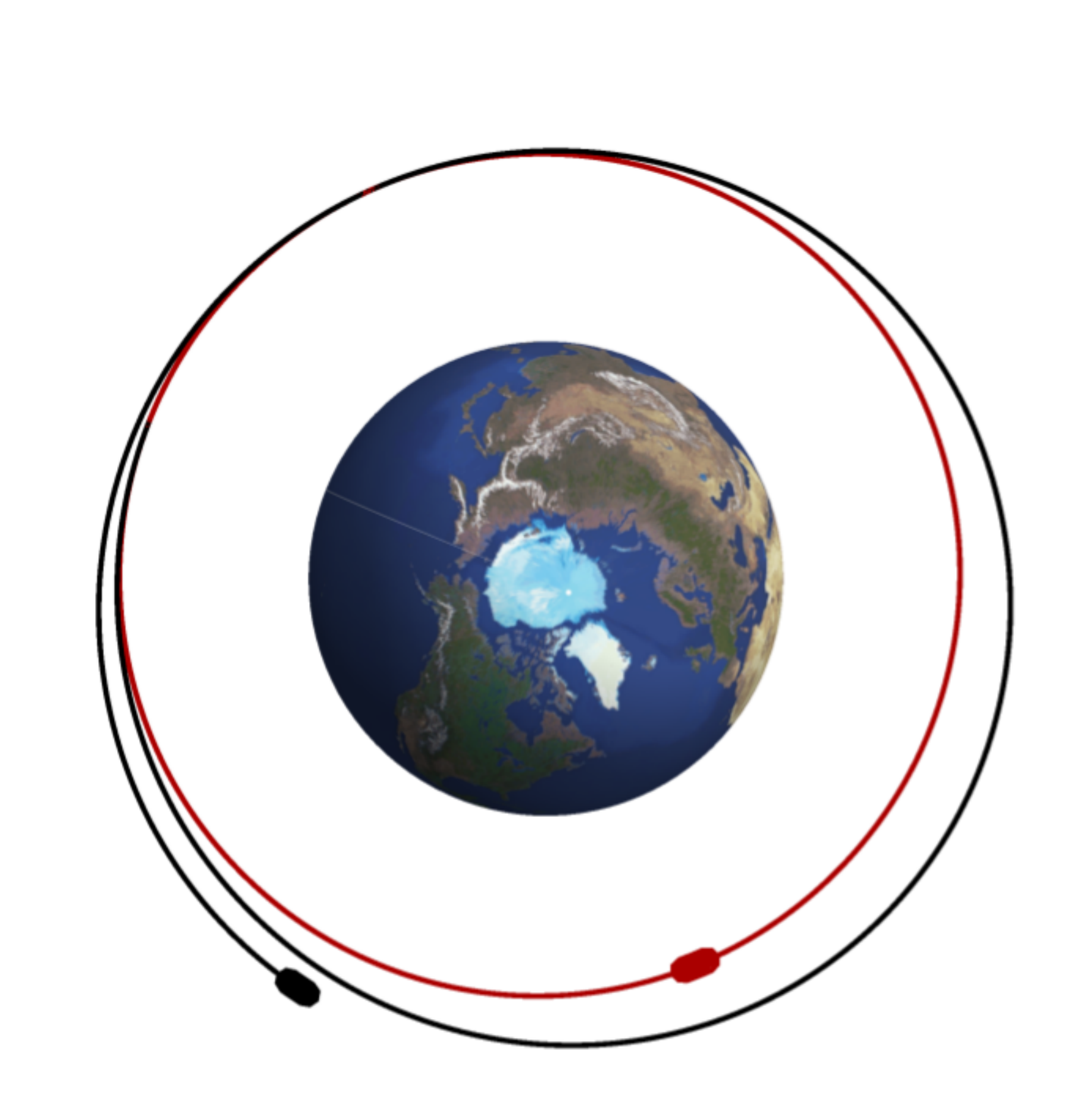}
\includegraphics[width=0.3\textwidth]{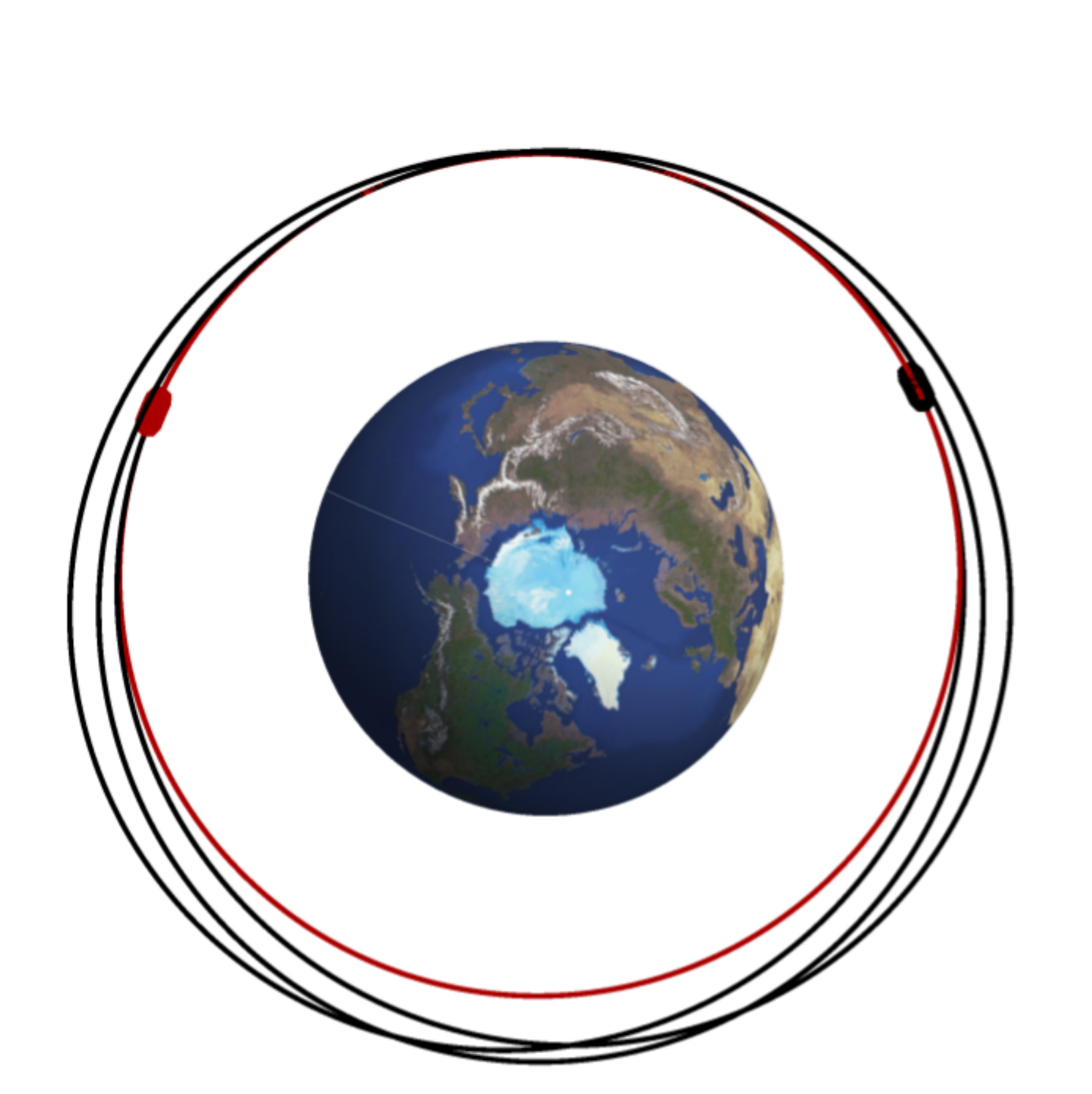}
\caption{\label{fig:EccOrbit} Solution of the Jacobi equation \eqref{eq:JacobiEqn} for satellite geodesics around the earth. The circular reference geodesic is shown in red and the deviating geodesic that corresponds to an elliptical orbit in black. The trajectories of both are shown for different values of the elapsed proper time on the reference orbit. The 3D orbits are shown in the top, while in the bottom row we display a top view to illustrate the eccentricity.}
\end{figure*}

In this section we show the results for deviation from a circular reference geodesic through an eccentricity. To construct such solutions of the Jacobi equation \eqref{eq:JacobiEqn} we choose the parameter according to
\begin{align}
C_1 = \text{arbitrary constant}\, , \quad C_3 = -2C_1 \dfrac{\sqrt{R_3}}{k^2} \, .
\end{align}
This choice ensures that both objects start from the same point in space and the radial components of the deviation vector $\eta^1(s)$ as well as its derivative $\mathrm{d} \eta^1/ \mathrm{d}s$ are initially (at $s=0$) equal to zero. $C_1$ will then determine the maximal possible radial deviation from the circular reference orbit. Note that this example was considered by Fuchs in \cite{Fuchs1983} as well, but there the choice of the constant $C_3$ has the wrong sign and there are several misprints at the indices of the constants. The perturbed orbit is given by
\begin{subequations}
\begin{align}
r(s) &= R_0 + \dfrac{E R_0}{\sqrt{L^2+R_0^2}} \dfrac{C_1}{k} \sqrt{\dfrac{4R_3}{k^2}} (1-\cos ks) \notag \\
&= R_0 + \delta r (1-\cos ks) \, , \\
\theta(s) &= \pi/2 \, ,\\
\phi(s) &= \omega s + C_1 \left( \left(1-\dfrac{4R_3}{k^2} \right) s + \dfrac{4R_3}{k^3} \sin ks \right) \, .
\end{align}
\end{subequations}
For a motion of that kind we derive the semi major axis $a$ that is simply given by
\begin{align}
a = R_0 + \delta r
\end{align}
and an eccentricity of
\begin{align}
e = \dfrac{\delta r}{R} \, .
\end{align}
The result is shown in figure (\ref{fig:EccOrbit}) for different values of the elapsed proper time. We can clearly see the effect of perigee precession that was examined already by Fuchs for this specific example \cite{Fuchs1990}. The $r$ and $\phi$-motions involve different frequencies, i.e., $\omega$ and $k$. For a full orbit that starts at radius $r(s=0)=R_0$ and ends at $r(s=2\pi/k) = R_0$ the elapsed proper time is $s = 2\pi/k$ and yields an azimuthal angle $\phi(s=2\pi/k) = \omega ~ 2\pi/k$. The difference to $2\pi$ is now called the perigee precession $\Delta \phi$. Thus, we get, see e.g. \cite{Fuchs1983},
\begin{align}
\Delta \phi = \dfrac{\omega ~ 2\pi}{k} -2\pi = 2\pi \left (\dfrac{\omega}{k} -1 \right) \, .
\end{align}
Inserting $\omega$ and $k$ that were given before in terms of the circular radius $R_0$ gives the known value
\begin{align}
\Delta \phi = 2\pi \left( \sqrt{\dfrac{R_0}{R_0-6M}} - 1 \right) \, .
\end{align}
However, in the Newtonian case there is no such quantity as $k$ and only the Keplerian frequency $\omega_0$ appears instead of $k$. Thus, for the Newtonian case we get no perigee precession and the behavior is qualitatively different in comparison to that shown in Fig. (\ref{fig:EccOrbit}) - the perturbed orbit, i.e., the ellipse just remains unchanged.
\section{Conclusion and Outlook}
Modeling satellite configurations by means of the Jacobi equation produces qualitatively good results in the sense that known effects are reproduced. However, it has to be carefully determined to which accuracy the results can be used compared with two direct solutions of the geodesic equation since the Jacobi equation was linearized w.r. to relative distance and velocity between the two objects. In an upcoming paper we will give results for the comparison between solutions of the Jacobi equation and the direct solution of the geodesics equation for simple spacetimes.
For higher accuracy the generalized Jacobi equation might be used to achieve better accuracy. If the distance between both satellites is small but the relative velocity is not, one might do the linearization only w.r. to the deviation itself and keep higher order terms in the derivative. If the relative velocity is small instead but the spatial deviation is not, one should do it vice versa.

For the future import steps will be to obtain solutions of the Jacobi equation in more realistic, but therefore more complicated spacetimes as models of the real earth. The considered Schwarzschild spacetime is the simplest model that does not include the rotation or even higher multipole moments of the earth. For these more realistic spacetimes the solution of the Jacobi equation will certainly need numerical integration or approximation methods. 
\section*{Acknowledgment}
The authors would like to thank Dirk P\"utzfeld for insightful discussions. The first author acknowledges financial support from the German Collaborative Research Center (SFB) 1128. Support from the Research Training Group \emph{Models of Gravity} is gratefully acknowledged. VP was financially supported during this work by Deutsche Forschungsgemeinschaft.

%

\end{document}